\documentclass[sigconf, nonacm, pdfa]{acmart}

\newcommand\vldbdoi{10.14778/3827998.3828024}
\newcommand\vldbpages{4169 - 4181}
\newcommand\vldbvolume{19}
\newcommand\vldbissue{12}
\newcommand\vldbyear{2026}
\newcommand\vldbauthors{\authors}
\newcommand\vldbtitle{\shorttitle} 
\newcommand\vldbavailabilityurl{}
\newcommand\vldbpagestyle{empty} 

\usepackage[a-2b,mathxmp]{pdfx}
\usepackage{balance}

\begin{document}
\title{Real-time SQL Plan Management in Oracle}

%%
%% The "author" command and its associated commands are used to define the authors and their affiliations.
\author{Sunil Chakkappen}
\affiliation{%
  \institution{Oracle America Inc.}
  \streetaddress{P.O. Box 1212}
  \city{Redwood City}
  \state{CA}
  \country{USA}
}
\email{sunil.chakkappen@oracle.com}

\author{Mohamed Ziauddin}
\affiliation{%
  \institution{Oracle America Inc.}
  \city{Redwood City}
  \state{CA}
  \country{USA}
}
\email {mohamed.ziauddin@oracle.com}

\author{Hong Su}
\affiliation{%
  \institution{Oracle America Inc.}
  \city{Redwood City}
  \state{CA}
  \country{USA}
}
\email {hong.su@oracle.com}

\author{Shreya Kunjibettu}
\affiliation{%
  \institution{Oracle America Inc.}
  \city{Redwood City}
  \state{CA}
  \country{USA}
}
\email{shreya.kunjibettu@oracle.com}

\author{Nigel Bayliss}
\affiliation{%
  \institution{Oracle Global Services Ltd.}
  \city{Thames Valley}
  \state{Berkshire}
  \country{England}
}
\email{nigel.bayliss@oracle.com}

\begin{abstract}
Consistent query performance is essential for mission critical database applications, yet SQL execution plans can change due to factors such as database upgrades, DML changes, new indexes, etc. While plan stability mechanisms such as stored outlines prevent regressions by freezing execution plans, they also inhibit performance improvements by disallowing plan evolution. We introduced SQL Plan Management (SPM) in Oracle 11g to address this trade-off by maintaining a set of accepted execution plans and allowing plan evolution only when new plans demonstrably outperform existing baselines. However, prior implementations of SPM primarily rely on background performance verification processes, delaying regression detection and recovery. This issue is amplified in autonomous cloud database systems, where several automatic actions that could cause plan change driven regressions are performed with limited customer control. Timely detection and remediation is paramount, but the constrained background resources on cloud may not keep pace.

To overcome these limitations, we introduce Real-Time SPM in Oracle 26ai, a novel extension of SPM that performs foreground verification of new execution plans during user query execution. Real-Time SPM leverages runtime session context to immediately validate plan changes, enabling rapid adoption of superior plans while promptly detecting and preventing regressions. This paper presents the architecture and design of Real-Time SPM - including technical challenges like reliably comparing performance of previous plans - and contrasts it with traditional background plan evolution. Our experiments highlight tangible benefits of Real-Time SPM - delivering immediate performance boost while preserving plan stability in both cloud-native and on-premise environments. Real-Time SPM is successfully deployed in Oracle production, laying the groundwork for more adaptive and user-responsive SQL plan management in enterprise RDBMS platforms.

\end{abstract}

\maketitle

\pagestyle{\vldbpagestyle}
\begingroup\small\noindent\raggedright\textbf{PVLDB Reference Format:}\\
\vldbauthors. \vldbtitle. PVLDB, \vldbvolume(\vldbissue): \vldbpages, \vldbyear.\\
\href{https://doi.org/\vldbdoi}{doi:\vldbdoi}
\endgroup
\begingroup
\renewcommand\thefootnote{}\footnote{\noindent
This work is licensed under the Creative Commons BY-NC-ND 4.0 International License. Visit \url{https://creativecommons.org/licenses/by-nc-nd/4.0/} to view a copy of this license. For any use beyond those covered by this license, obtain permission by emailing \href{mailto:info@vldb.org}{info@vldb.org}. Copyright is held by the owner/author(s). Publication rights licensed to the VLDB Endowment. \\
\raggedright Proceedings of the VLDB Endowment, Vol. \vldbvolume, No. \vldbissue\ %
ISSN 2150-8097. \\
\href{https://doi.org/\vldbdoi}{doi:\vldbdoi} \\
}\addtocounter{footnote}{-1}\endgroup

\ifdefempty{\vldbavailabilityurl}{}{
\vspace{.3cm}
\begingroup\small\noindent\raggedright\textbf{PVLDB Artifact Availability:}\\
The source code, data, and/or other artifacts have been made available at \url{\vldbavailabilityurl}.
\endgroup
}

\section{Introduction}

Oracle-specific SQL Plan Management (SPM) ~\cite{SPM_blog} is a preventative mechanism and one of the multi-pronged approaches against performance regressions caused by plan changes. Regressions can be caused by many different reasons like environmental changes, bugs in the software updates, or stale statistics. While there have been numerous robust and adaptive query optimizations proposed ~\cite{Rome, Polar, par2q0, databtricks} - citing cardinality estimation errors as the primary cause of degraded execution plans ~\cite{cardinality} - these mitigation strategies, like any approach, do not guarantee zero regressions. They also \textit{reactively} try to correct a plan, or parts of it, when it is found to have degraded performance ~\cite{SPM}. Once a regression occurs, it has to be managed, and that is where SPM steps in. SPM addresses plan regressions in a \textit{proactive} manner ~\cite{SPM}, enabling the RDBMS Optimizer to automatically manage execution plans, ensuring that the database only uses plans verified to be better. It has three main components:
\begin{itemize}
\item Plan baseline capture - captures and stores relevant information about plans, called SQL plan baselines, for a set of SQL statements. 
\item Plan baseline evolution - verifies the performance of existing SQL plan baselines of a statement and marks them as accepted if performance is better compared to other plans for the statement.
\item Plan baseline selection - selects one of the accepted plan baselines for the statement and avoids potential performance regressions.
\end{itemize}

SPM was introduced in Oracle RDBMS 11g ~\cite{SPM, SPM_maria}. The whole cycle of plan management was mostly manual, starting with the user loading the SQL statements and their plans into SPM repository. Only plans accepted by the user are considered for execution. Whenever a brand new plan is generated by Cost Based Optimizer (CBO), it is added into SPM repository as a non-accepted plan. The user has to use manual system APIs to evolve a newly added plan by verifying its performance against the baseline performance. If a new plan is accepted, then the CBO starts considering that plan upon next compilation. 
 
In Oracle 12c, Automatic SPM Evolve Advisor ~\cite{12c} was introduced, shifting away from manual work of evolving plan baselines. With this, the verification and evolution of existing plan baselines was automated as a background task triggered by the existing "SQL Tuning Advisor" ~\cite{Tuning} client under the automated database maintenance task framework. This moved the burden of plan performance verification and baseline evolution of existing plan baselines from the user to the system. However, users still had to handle the regressing SQL statements by capturing good plans as SQL Plan Baselines using manual APIs.
 
Database developers and administrators (DBAs) may be faced with a daunting task dealing with plan changes and regressions resulting from database upgrades/patching or data modifications/new access path structures like indexes in the tables. Since controlled and fully automated plan management systems will be extremely useful in such cases, Automatic SQL Plan Management (Auto SPM) ~\cite{19c} was introduced in Oracle 19c with the goal to provide an automated end-to-end solution to detect and correct regressions for autonomous Exadata environments. Auto SPM relies on another infrastructure component implemented in RDBMS 19 called the Automatic SQL Tuning Set (Auto STS) ~\cite{STS}. Auto STS, described further in section 2.3, periodically captures statements executed in the system and their performance metrics via a background task. Auto SPM was implemented using a background task which runs periodically for up to an internally defined amount of time in which it captures plans as unverified from Auto STS into the SPM repository. 
Capturing plans into the SPM repository consists of:
\begin{enumerate}
\item looking for multiple plans in Auto STS for a SQL statement
\item comparing performance metrics (a function of CPU time and Buffer Gets) of best and worst plan
\item if the best plan performance benefit is above the margin (“regression threshold”) then capture all the plans except for the worst one as non-accepted plan baselines
\end{enumerate}
Auto SPM captures plans for any query that exemplifies significant performance variations from its historical plans. It then verifies the performance of unverified/non-accepted plans, thus creating new accepted plan baselines by evolving non-accepted plans.

However, reliance on background execution introduces several limitations for Automatic SPM, as further discussed in Section 2.4. Background verification was a natural choice for on-premise database systems, but proved to be insufficient for database on cloud. In cloud-managed services, software patches are applied periodically and automatically, often without direct customer involvement. Other automatic actions like Oracle's Automatic Indexing ~\cite{Autoindex} or Automatic Optimizer Statistics Gathering ~\cite{autostats} are also performed in autonomous databases. Such actions can introduce plan regressions, yet background verification tasks may not react quickly enough to prevent user-facing performance impact, particularly under constrained resource budgets when resources are shared among different tenants. Background processing also incurs additional resource consumption, increasing operational cost for customers - both on premise and on Cloud. In elastic cloud settings, rapid fluctuations in resource demand can further amplify cost and latency instability in analytical workloads, even under ideal scaling policies ~\cite{cackle}. Since background verification is asynchronous with foreground query execution, regression detection and prevention are inherently delayed. Due to the reasons above, such limitations are more prominent on a cloud database compared to the traditional on-premise model.

\begin{figure}[h]
  \centering
  \includegraphics[width=\linewidth]{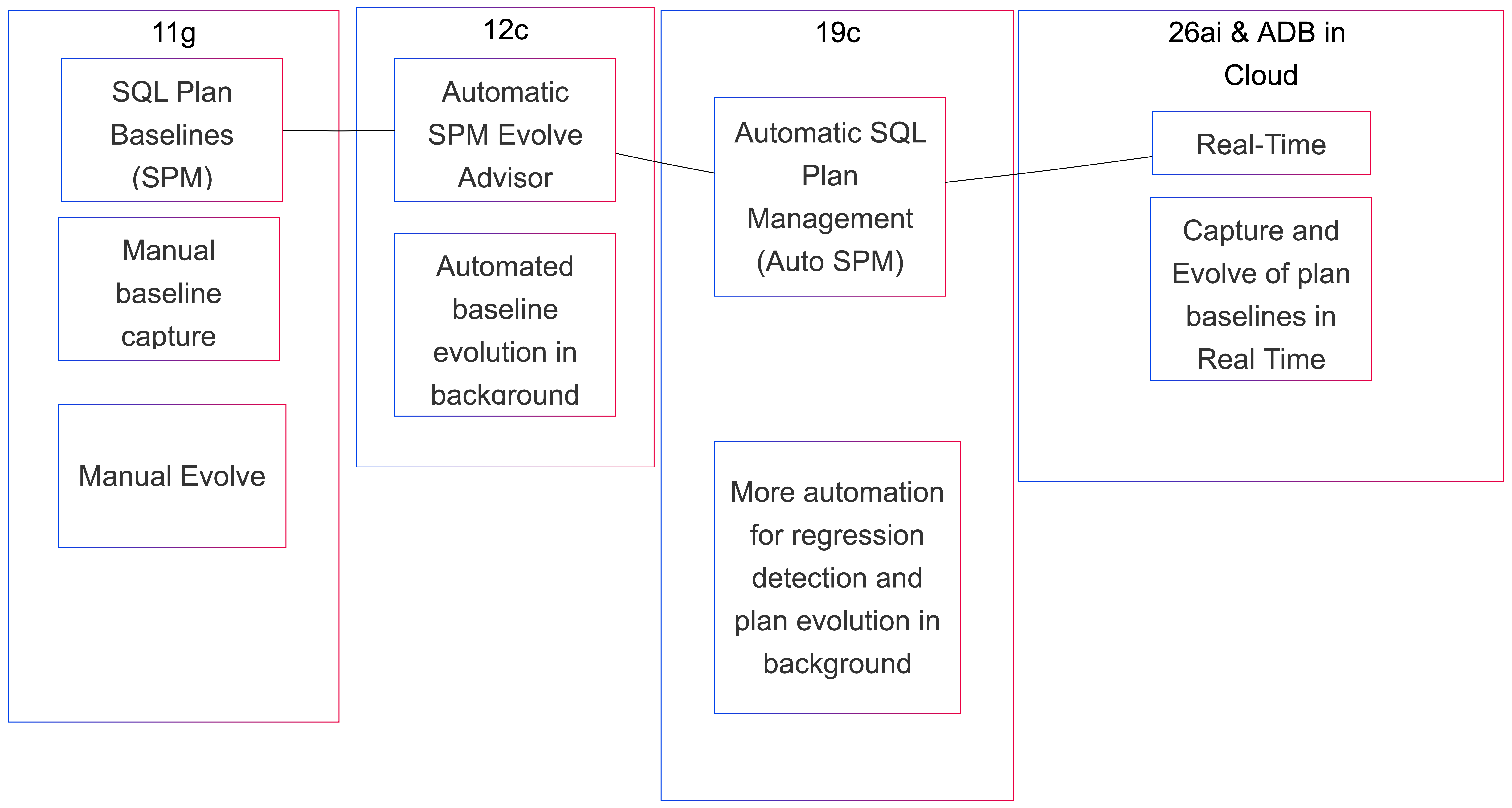}
  \caption{SPM Timeline}
  \label{fig:timeline}
\end{figure}

We introduce Real-Time SPM with the intent to reduce resource usage in the system, and enable automatic SQL plan management on-Cloud on fully managed services in an effective way. A CBO plan that has not yet been accepted can execute in the foreground and have its performance verified during user execution, rather than being deferred to background verification. While newly generated CBO plans are often superior to previously observed plans, if any query exhibits performance regressions, Real-Time SPM reinstates a previously accepted plan and guides the optimizer to avoid the regressed plan in subsequent executions. If the newly generated CBO plan is better in performance, the CBO plan will be used going forward. This approach enables faster detection and resolution of plan regressions, ensures that users benefit from high-performing plans as early as possible, and reduces reliance on background processing by capitalizing on performance metrics collected during Real-Time query execution. \autoref{fig:timeline} summarizes the progression of SPM functionality over Oracle releases.

Section 2 describes Auto SPM in Oracle 19c and its limitations in more detail, followed by the Real-Time SPM solution details in section 3. We provide our experimental results and comparison of Real Time SPM vs Auto SPM in section 4, tying it in with Related Works in section 5 and concluding in section 6. 

\section{Auto SPM}
SPM operates in both foreground and background. A statement is managed by/under SPM if 
it has at least one plan baseline captured into the SPM repository. In the foreground, SPM captures new plan baselines for statements already under SPM. In the background, it detects regressions and new statements to be managed by SPM - capturing and evolving new plan baselines for them. Each module is described below in detail.

\subsection{Auto SPM Foreground Actions}
In the foreground, SPM is responsible for capturing new CBO generated plans for statements under SPM. As soon as a plan is generated by the CBO, a fast in-memory bit vector check is done to check if the current statement is managed by SPM. This bit vector is on SQL text signature, where a bit is set for each statement under SPM or if it has SQL Management Objects (SMO) baselines. If the bit is set, it further checks if plan baselines exist for the statement by loading them into the cache. If the statement is not under SPM, then the CBO generated plan is used for the execution and the statement remains as NOT under SPM.

Upon passing the bit vector check, basic information (plan hash value and plan baseline status) for all plans associated with the text signature of the current SQL statement is loaded into the cache. If the CBO plan matches with an accepted plan, then it is used for the execution. 

\begin{figure}[h]
  \centering
  \includegraphics[width=\linewidth]{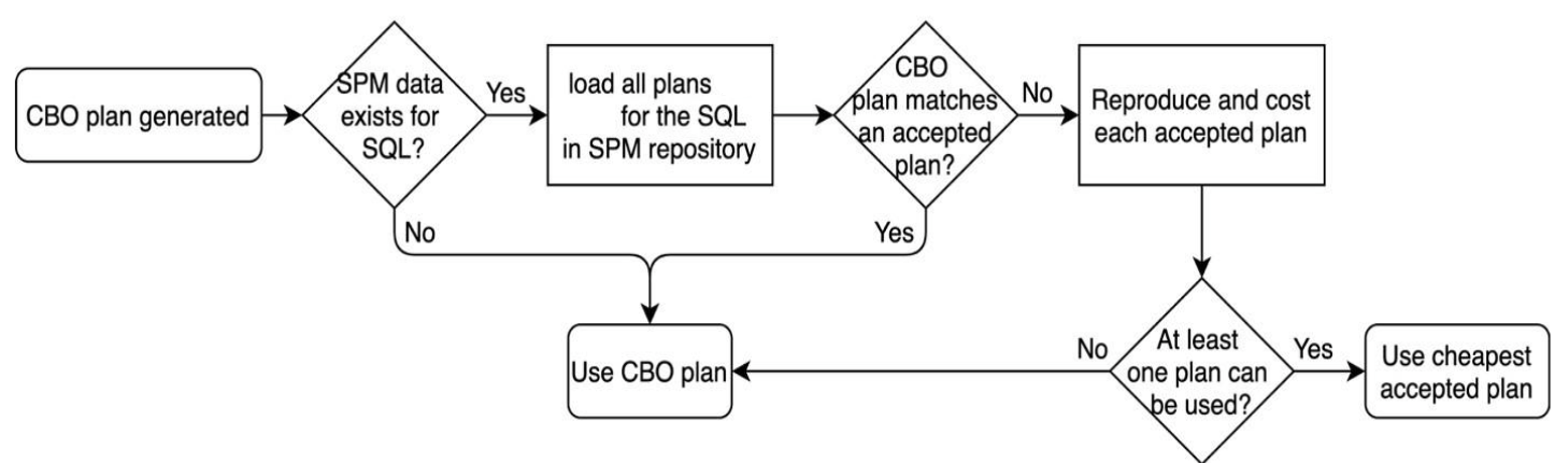}
  \caption{Foreground Plan Selection}
  \label{fig:foreground}
\end{figure}

If the CBO plan is a new or non-accepted plan, i.e. previously seen but not verified to be better, then more information, such as outline data containing set of optimizer hints that encode the key plan choices for a previously observed plan, is loaded for all the accepted plans. Each accepted plan baseline is then reproduced using outlines by recompiling the SQL statement while applying those stored hints in an attempt to regenerate the same execution plan structure and costed in the user environment with current binds. The plan with the least optimizer cost among all the reproducible accepted plans is used for the execution. If none of the accepted plans is reproducible then the CBO plan is used for the execution and captured into SPM as a non-accepted plan. \autoref{fig:foreground} depicts the foreground SPM processing flow.

\subsection{Auto SPM Background Actions}
\begin{figure}[h]
  \centering
  \includegraphics[width=\linewidth]{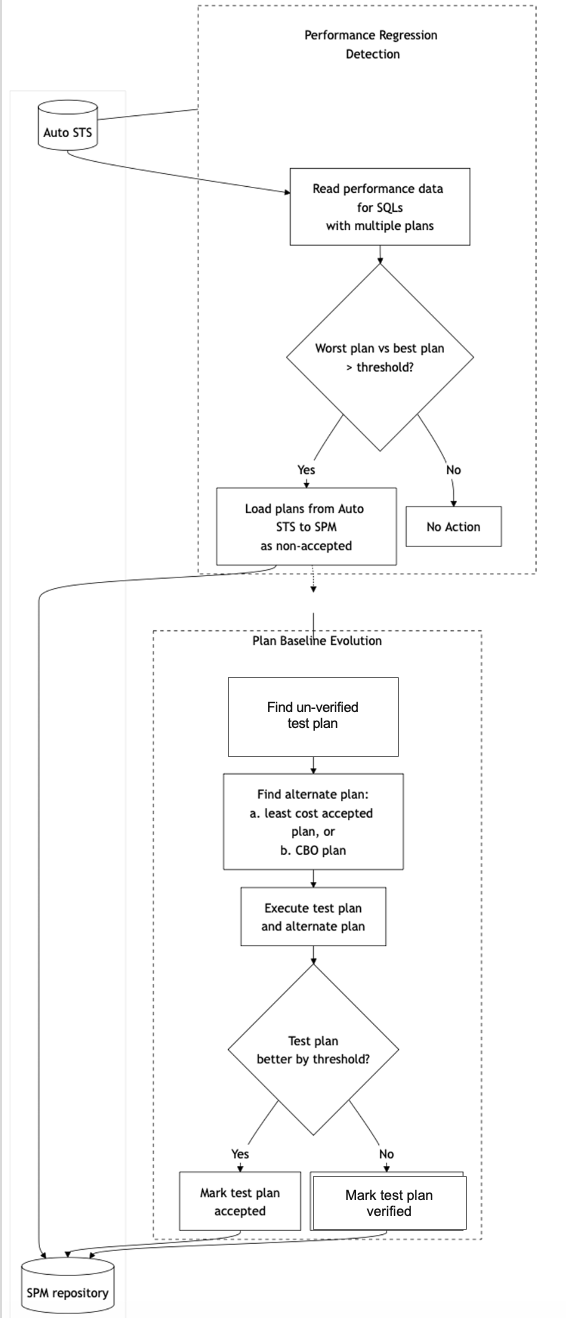}
  \caption{Background SPM}
  \label{fig:evolve}
\end{figure}

In the background, SPM detects and fixes regressions for the statements currently not managed by the SPM. It also verifies and evolves non-accepted plans for statements under SPM. Pictured in \autoref{fig:evolve}, background SPM processing is divided into two sub-modules - Performance Regression Detection and Plan Baseline Evolution.

\subsubsection{Performance Regression Detection}
The regression detection module detects and adds alternate plans into the SPM repository from Auto STS for potentially regressed statements. These plans are later verified in the Plan Baseline Evolution module and their statuses are set (i.e. accepted or non-accepted) based on their performance. A new plan is only captured into the SPM repository if it does not already exist there.

The potentially regressed statements detection starts by reading historical performance data for all statements from Auto STS that have historically used multiple plans. It then compares the historical performance (a function of CPU time and Buffer Gets) of the worst plan to that of the best plan. If the best plan shows significant performance improvement over the worst plan, then all the plans except for the worst plan are added into the SPM repository. If historical performance of the best and worst plan is less than an internally-defined Auto STS margin, it means that all the plans have historically shown similar performance. Using one plan over another through SPM is unlikely to show significant performance improvement in this case. Furthermore, as the task consumes valuable user resources in the background verifying performance and evolving the plans, filtering out statements with plans with similar performance facilitates efficient allocation of user resources. 

\subsubsection{Plan Baseline Evolution}
Plan baseline evolution, or the evolve driver, has been part of SPM since 11g as mentioned in section 1. SPM repository can have plans with following statuses:
\begin{enumerate}
 \item ACCEPTED, VERIFIED - Performance is better than some other plan and evolve accepted it. 
 \item NON-ACCEPTED, VERIFIED - Performance is worse/similar than some other plan and evolve did not accept it. 
 \item ACCEPTED, UNVERIFIED - User manually added the plan so it was accepted. 
 \item NON-ACCEPTED, UNVERIFIED - A new CBO plan captured into SPM will have this status. 
\end{enumerate}
The evolve driver reads the data from the SPM repository for all the plans that are either non-accepted and unverified or accepted and verified more than an internal period of days ago. It also avoids capturing plans for old statements, i.e. statements executed more than another internal retention period of days ago, as the SQL details may not be available past the retention period, and SPM tries to identify only important statements from \textit{current/recent} activity. It verifies each non-accepted plan by choosing an alternate plan to compare its performance against and then test executes them. The alternate plan is an accepted plan from the SPM repository with the least optimizer cost computed with the settings and binds of the unverified plan. If there is no reproducible accepted plan to compare the non-accepted plan against, then the CBO plan is used as the alternate plan and is executed alongside the non-accepted test plan. If the non-accepted plan is same as the CBO plan, then verification is skipped for that plan. 

Upon completion of the test execute, the non-accepted plan and the alternate plan performances are compared. If the non-accepted test plan is better by an internally-defined margin, then the test plan's status is modified to ACCEPTED, VERIFIED. Otherwise, the plan remains non-accepted but marked verified.

\subsection{Auto STS}
Auto SPM relies on Auto STS to identify repeatable statements and their historical performance. SQL tuning set (STS) ~\cite{STS} is a general purpose named database object that can store one or more SQL statements, along with their plans, execution metrics and execution context. Each Oracle database has an automatically created STS called Auto STS created during database creation and used by many automated SQL tuning features, including automatic SPM. SQL statements can be captured into an STS from different sources; for Auto STS, it periodically captures the user’s workload from shared cursor cache (i.e. cached execution plans). For any statement already existing in the STS, all the performance metrics in STS will be replaced with the new values from shared cursor cache. Any statement in STS not executed within a certain retention period will be purged ~\cite{Autoindex}.

\subsection{Auto SPM Limitations}

Despite the automated capabilities of SPM, being run in the background subjects it to the following shortcomings, including unfavorability in Cloud environments and asynchronous performance verification with foreground processes as referenced in section 1:
\begin{itemize}
\item May fail to reproduce plans and their actual performance. Being run in the background also implies that the SPM task may fail to exactly replicate the user environment on which the plan was generated, rendering it difficult to verify the plans. This might be due to failure to capture some environment settings, or due to bugs in the outline that fail to reproduce the specific plan. Virtual Private Database (VPD) predicates ~\cite{Vpd} - Oracle's security feature that enforces row-level and column-level security by automatically modifying SQL statements at compile time - further exacerbate this problem, as they might get applied in the user environment but not in the background task environment. 
\item Doesn't look at all statements (to reduce resource consumption). 
\item Auto SPM Task background process runs with limited resources.
Resources (CPU, Degree of Parallelism, time, etc.) are quite restricted for this background task, therefore increasing its processing time and adding more time windows between regression detection and regression resolution -  keeping customers waiting to see benefits of new good plans or prevention of regressed plans.
\item Fails to verify long running statements.
The Auto SPM Evolve task is allowed to run only for a short amount of time, like 30 minutes. It is not enough time to verify performance of long running test/reference plans. The Auto SPM Task running with limited resources further compounds this problem, increasing the chances of not verifying long running statements in the allotted window.
\end{itemize}

\section{Real-Time SPM}
As most of these challenges stem from verification occurring in the background, we eliminate the source of the shortcomings through Real-Time SPM, where verification now occurs in the foreground as well.
Two types of plans - test and reference - will be used frequently when describing Real-Time SPM. A \textit{test} plan is a CBO generated, NON-ACCEPTED, UNVERIFIED plan. There is another different plan, a \textit{reference} plan, loaded from the SPM repository or Auto STS that can be used for comparing performance. This reference plan is the optimizer-computed lowest-cost ACCEPTED, VERIFIED plan from the SPM repository that can be reproduced, or the lowest-cost reproduced plan from Auto STS.

Plan baseline selection is enhanced in that we can identify a new test plan and execute it once for performance comparison in real-time. We can also pick a relevant historical reference plan to compare test plan performance. Plan baseline evolution is also enhanced and performed in the user session. We collect performance metrics (like CPU time and buffer gets) at the end of test plan execution and compare the test plan metrics with average performance metrics of the reference plan:
\begin{itemize}
\item If test plan performance is better, ACCEPT test plan in SPM repository.
\item If test plan is worse, ACCEPT reference plan if it is not already, and test plan is avoided going forward.
\item If performance is similar, no action is taken.
\end{itemize}
In this way, we avoid the limitations of background auto SPM, and are also able to mitigate performance regression with only a \textit{single} execution of a potentially suboptimal (test) plan.

There are a couple challenges posed with having everything run in real-time. Since all the work is done when the user compiles/executes a statement, it can introduce an overhead of finding and comparing a reference plan in the foreground. Performance metrics of reference plans are also historical. We need an efficient strategy for cases when the historical performance of a reference plan does not represent the current performance had it been executed now. The most innovative and challenging aspect of Real-Time SPM is to compare the performance of a test plan to the historical performance of a reference plan in a reliable manner. The following sub-sections describe how we overcame these challenges along with detailing all aspects of Real-Time SPM.

\subsection{Foreground Plan Selection and Verification}
Real-Time SPM may proactively create new plan baselines or evolve existing plan baselines by executing the test plans instead of using the accepted plans from SPM repository. It uses the user session context for active/real-time performance verification rather than depending on the background doing passive/delayed performance verification. When a CBO-generated plan is detected as a test plan along with a reference plan chosen from the SPM repository or the Auto STS, then foreground performance verification in the user session is performed by executing the CBO plan. A CBO plan cannot be a test plan if there is no reference plan. In this case, the Auto SPM processing described in section 2.1 takes place. 

When selecting a reference plan in Real-Time SPM, if no reproducible accepted plan baseline exists, then the least-cost plan (based on optimizer cost using current binds and optimizer environment) is chosen from Auto STS as the reference plan. In the old background SPM (section 2.2), test plan is one of the unverified plan and reference plan will be CBO plan if there is no accepted plan. This difference is because in foreground the test plan is the CBO plan itself. Since we need a benchmark to compare the CBO plan performance, the least cost Auto STS plan is used as the reference plan. 
 
After execution, the collected test-plan performance metrics are compared against the stored performance metrics of the reference plan, and the better-performing plan is ACCEPTED (i.e., either the test plan or the reference plan, if not already accepted). A test plan that is verified but not accepted is marked for reverse verification so that if the CBO generates that plan again in a subsequent compilation, reverse performance verification can be performed as explained in section 3.5. 

\begin{figure}[h]
  \centering
  \includegraphics[width=\linewidth]{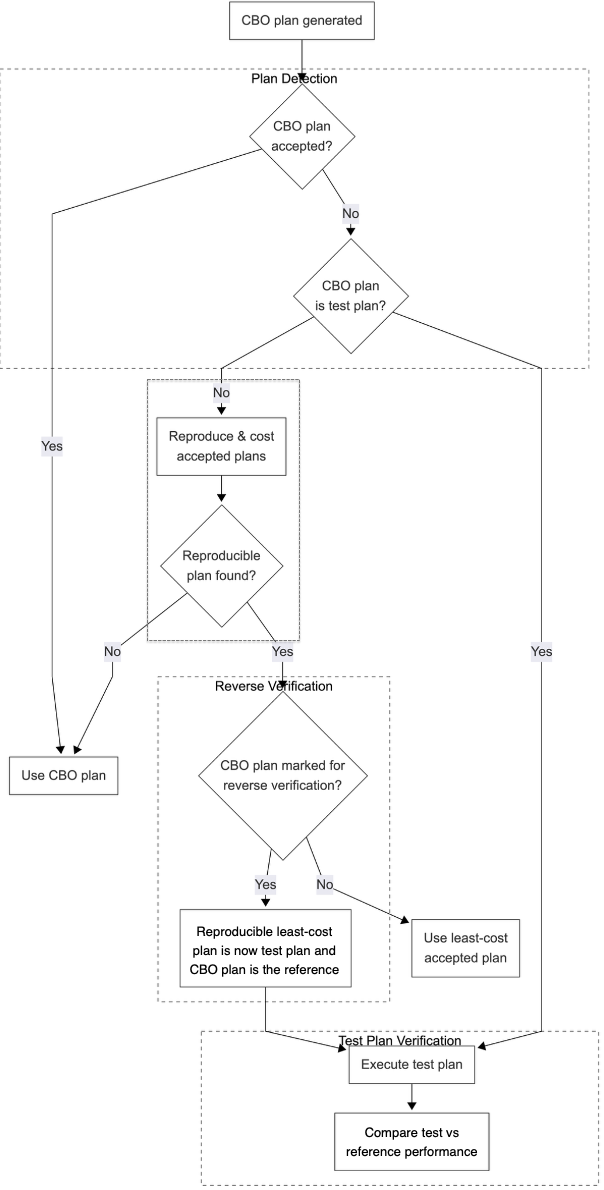}
  \caption{Real-Time SPM}
  \label{fig:newforeground}
\end{figure}

Real-Time SPM is divided into four modules - test plan detection, reference plan detection, test plan performance verification, and reverse verification, each of which is described in the following subsections. \autoref{fig:newforeground} captures how these four sub-modules work together, with each sub-module pictured inside dotted boxes. Section 3.6 contains examples to help walkthrough \autoref{fig:newforeground}.

\subsection{Test Plan Detection}
When the optimizer generates a CBO plan, Real-Time SPM must determine whether that plan should be treated as a test plan for foreground verification. A CBO plan qualifies as a test plan only if both of the following hold:
\begin{enumerate}
    \item the plan has not already been foreground verified, and
    \item at least one different reproducible historical plan exists that can serve as a reference
\end{enumerate}
The first condition prevents repeated verification of the same plan. The second ensures that verification is meaningful, since a test plan can only be evaluated if there is an alternative plan for comparison.

To check the first condition, Real-Time SPM first determines whether plan baselines already exist for the statement. If none exist, then the CBO plan has not been verified before. If baselines do exist, they are loaded from the SPM repository into the cache, and the CBO plan is checked against the existing accepted and previously verified plans as shown in \autoref{fig:foreground}.

To check the second condition efficiently, Real-Time SPM uses two bit vectors derived from Auto STS. Querying Auto STS directly during compilation would add too much overhead, especially because most statements have only one historical plan. The first bit vector indicates whether a statement has multiple plans in Auto STS. If this bit is set, then a plan different from the current CBO plan must exist, so the CBO plan is a valid test-plan candidate.

If the multi-plan bit is not set, then the statement has either one historical plan or none. In that case, Real-Time SPM consults a second bit vector that records whether the current (SQL ID, plan hash value [PHV]) pair exists in Auto STS. If it does, then the CBO plan is the only historical plan and no different reference plan exists, so foreground (FG) verification is skipped. Otherwise, either a different plan exists or the statement is still new to Auto STS, and Real-Time SPM proceeds to reference-plan detection.

\begin{figure}[h]
  \centering
  \includegraphics[width=\linewidth]{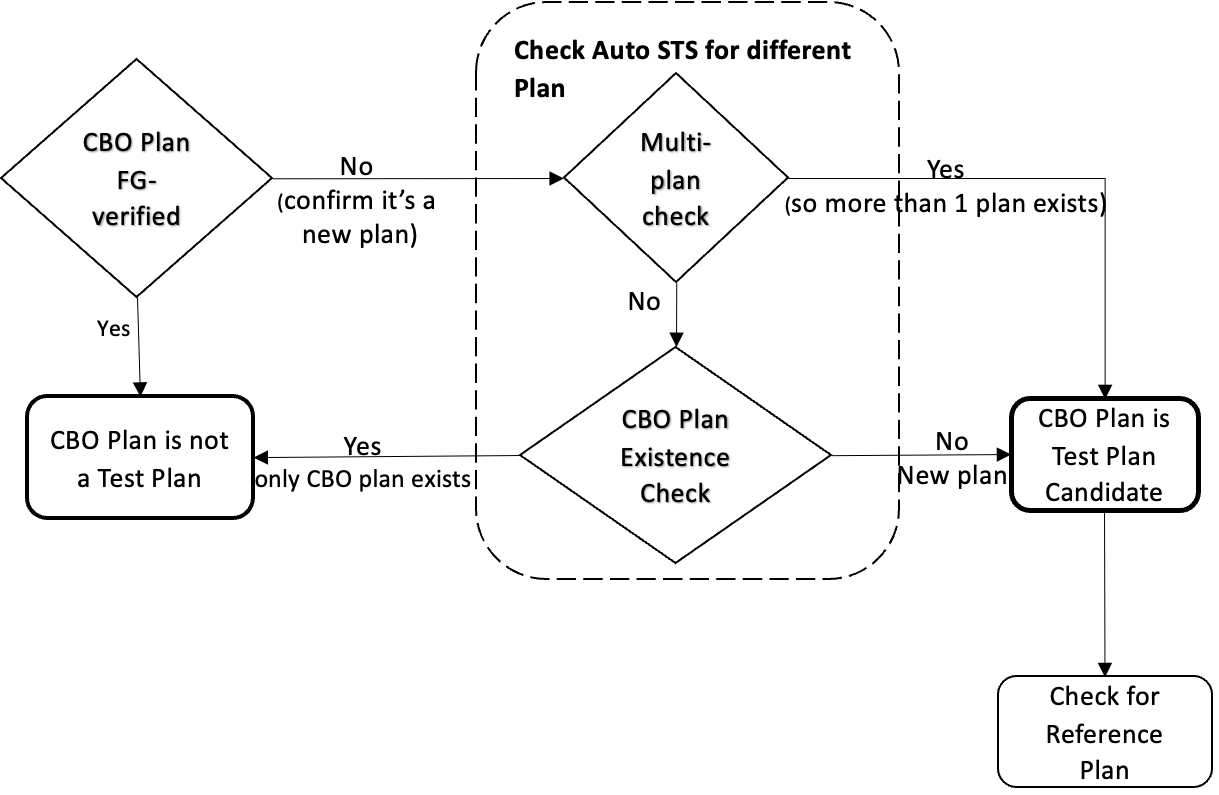}
  \caption{Test Plan Detection}
  \label{fig:tpd}
\end{figure}

In this way, test-plan detection avoids unnecessary Auto STS lookups for the common case where there is a single plan for a statement, while still identifying statements for which foreground verification is possible. \autoref{fig:tpd} outlines the different plans' detection methodology.

\subsection{Reference Plan Detection}
Once the CBO plan has been determined as a test plan candidate, foreground verification will proceed to identify the reference plan to compare against the test plan. As established, a reference plan is one of the historical plans that has execution metrics and is reproducible. 

SPM repository will first be probed for reproducible accepted plans. If such plans exist, then verification will retrieve associated plan metrics like optimizer cost and execution statistics from Auto STS for plan comparison to verify if test plan is better. If metrics are found for some plans, those plans are compiled and costed by CBO. The least cost plan is selected as the reference plan. 

\begin{figure}[h]
  \centering
  \includegraphics[width=\linewidth]{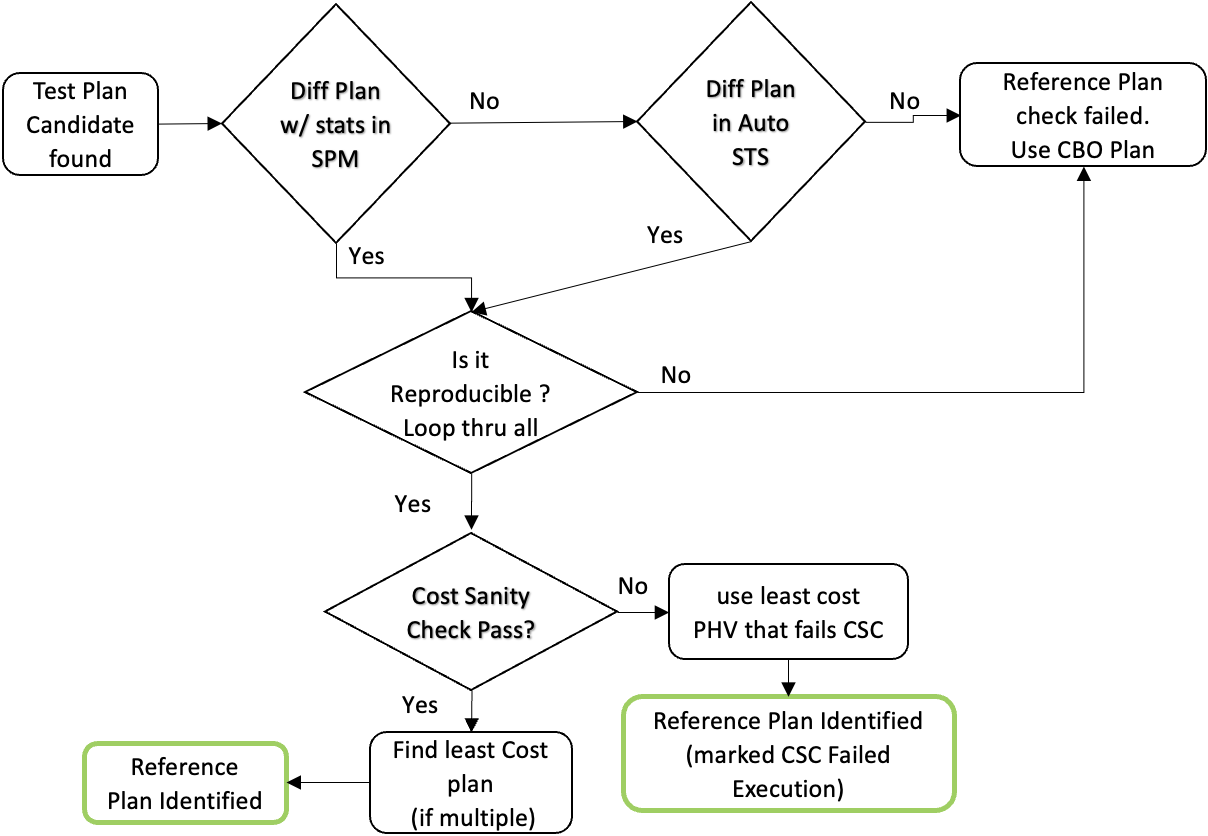}
  \caption{Reference Plan Detection}
  \label{fig:refPlan}
\end{figure}

Accepted plans may fail to reproduce due to changes in schema metadata, such as dropped indexes, or user-session context, such as new VPD predicates ~\cite{Vpd}. If no reproducible accepted plans from SPM with performance metrics are found, then Auto STS is scouted for a reference plan. All the non-accepted Auto STS plans other than the test plan are compiled and costed by the CBO to pick the least cost plan as the reference plan. If none of the Auto STS plans are reproducible either, then verification will be skipped as no reference plan was detected. \autoref{fig:refPlan} shows the reference plan detection steps, first looking in SPM repository and then within Auto STS.

 Another challenge of Real-Time SPM is that historical performance might not represent current performance of the same plan. To overcome this, we introduce cost-sanity checks that a plan must pass to be considered a reference plan. The historical optimizer cost of the plan should be close to the cost computed now based on current binds and compilation environment. Refer to section 3.6.1 for more details on bind-sensitive plan choice.
 
 The cost-sanity checks are built-in to the least-cost plan computation. The current plan is a valid reference plan if:
\begin{enumerate}
\item Its cost in the current environment has not gone up or down a lot, e.g. due to increase or decrease in number of rows in some tables. The new cost to old cost ratio, or vice-versa, should be less than an internally-defined margin.
\item The difference between the new cost and the old cost of the current plan is less than an internally-defined margin. For plans with small cost, the ratio in item 1 above can be very high, but we consider it for reference plan selection if the absolute difference is not high. These margins are fixed, and derived from the same internal improvement margins as in sections 2.2 and 3.4.2.
\end{enumerate}

However, even if the best-available reference plan failed cost-sanity checks, it cannot be assumed that the test plan is better in the current environment. In the event that we do not find a reference plan that passes cost-sanity checks, we still use the least-cost plan as a reference plan. During Test plan performance verification, we will not accept the test plan even if it is better since we do not fully trust the performance of the reference plan in the current environment. We are more conservative in accepting the brand new test plan. However, the reference plan - a plan seen in the past - will get accepted if it is better than the test plan and will go through reverse verification (described in section 3.5) in later executions.

\subsection{Test Plan Performance Verification}
 This module executes and compares the performance of the test plan against average historical reference plan metrics and accepts the winner to maintain the current performance benefit or to avoid current performance regression going forward. It is made up of two parts, test plan execution and plan performance comparison, each of which is described below.

 \subsubsection{Test Plan Execution}
 Upon selection of the reference plan, the CBO executes the test plan. In the product setting, Real-Time SPM does not impose a runtime cutoff on a test-plan execution, since doing so could introduce user-visible failures. The design is intended to limit the impact of a bad plan to at most one foreground execution, but not to bound the worst-case duration of that execution. Test plan performance metrics will be collected and compared against the historical reference plan average performance metrics in the plan comparison sub-module.

 \subsubsection{Plan Performance Comparison}
 The possible situations are:
 \begin{itemize}
     \item Test plan is better. If the test plan performance is better than the average reference plan performance by an internally-defined margin, then the test plan status is changed to ACCEPTED and VERIFIED as long as the test plan execution is NOT interrupted or the reference plan cost sanity check mentioned in section 3.3 has NOT failed. If the test plan is interrupted for some reason (e.g. due to resource manager timeouts or customer interrupts), we do not know the complete performance metrics of the test plan and hence can not accept it. 
     \item Test and reference plan performance is similar. In this case, the test plan is marked verified.
     \item Test plan is worse. If the performance of the test plan is worse than average reference plan performance by an internally-defined margin, then the reference plan will be ACCEPTED if it was not already, and test plan will be marked VERIFIED. Note that we can do this even if the test plan was interrupted as the metrics are worse even with partial execution. If the reference plan is accepted, the test plan will be marked for reverse verification as explained in section 3.5. SPM never rejects or blacklists a test plan based on one regressing execution, since that same plan may still be beneficial in a different execution context. Instead, it avoids the non-accepted test plan until its performance is shown to be better through normal or reverse verification.
 \end{itemize}

  \begin{figure}[h]
  \centering
  \includegraphics[width=\linewidth]{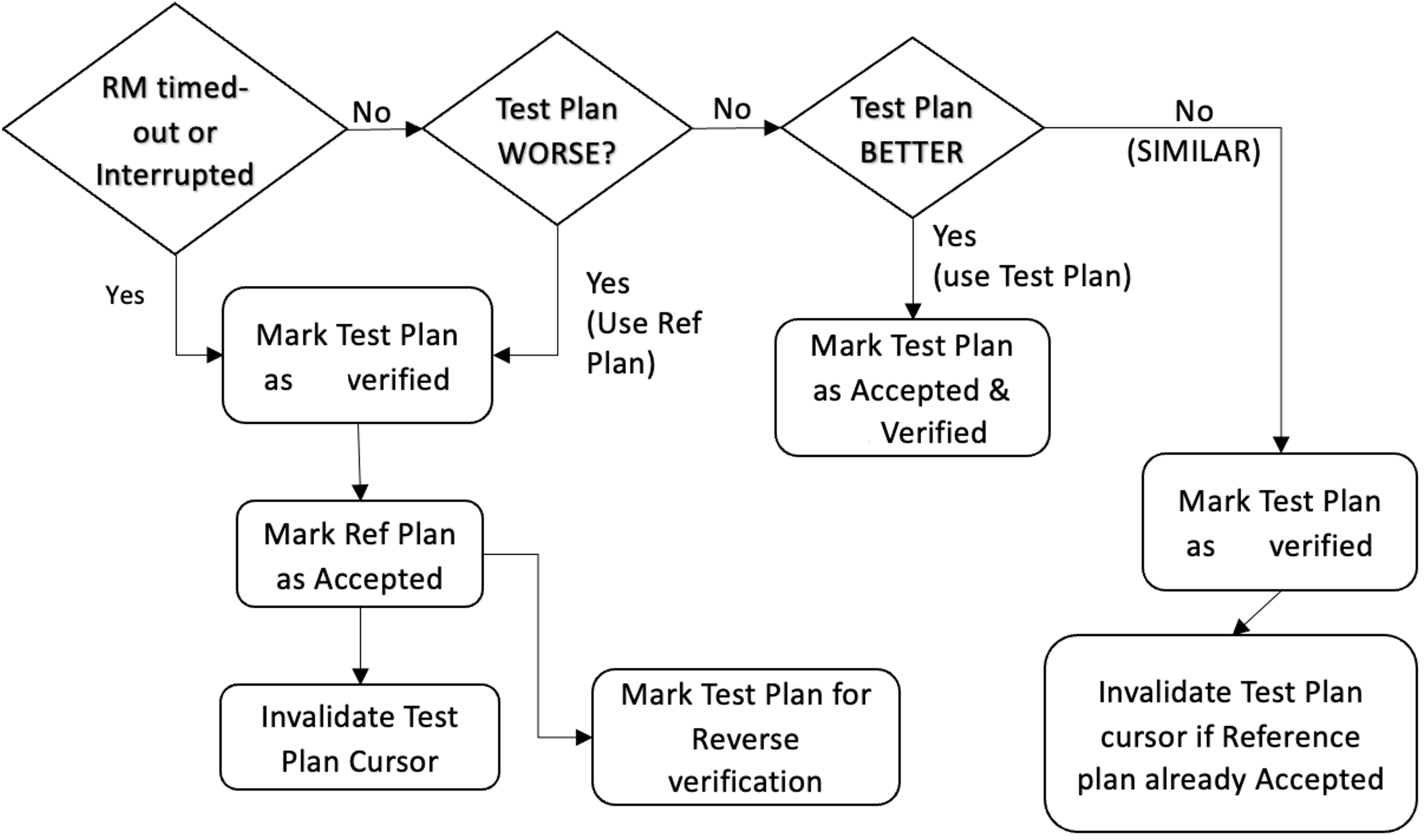}
  \caption{Test Plan Verification}
  \label{fig:planVerification}
\end{figure}

 \autoref{fig:planVerification} depicts the test plan verification workflow. Real-Time plan performance verification helps us avoid repeated execution of a bad test plan, in contrast to repeated bad executions until performance verification by Auto SPM in background.

The test plan’s cursor is also invalidated if its performance is deemed similar to or worse than that of the reference plan. Oracle stores optimizer decisions and execution plans in a structure called a cursor~\cite{Autoindex}. Cursors are cached in a shared-memory area called the cursor cache, allowing repeated executions of the same SQL statement to reuse an existing compiled plan. If the reference plan is similar to or better than the test plan, the test-plan cursor is invalidated; otherwise, it could be reused on the next execution. Even when performance is similar, Real-Time SPM conservatively prefers previously accepted plans unless a new plan shows a clear improvement. Invalidation therefore forces the next execution back through plan selection, where it can fall back to a reproducible accepted plan.

\subsection{Reverse Verification}
Real-Time SPM uses a combination of cost sanity check and reverse plan verification to overcome the challenge of comparing test and reference plans in a reliable manner. A test plan is marked for reverse verification only if performance is worse and a reference plan is accepted. For cost sanity check-failed cases, test plans marked will be reverse verified in subsequent executions. More intuition behind reverse verification can be found in section 3.6.

If the average historical performance of the reference plan is better compared to the test plan and reference plan was a non-accepted plan, the reference plan will be accepted. The test plan will have foreground verified status and will be marked for reverse verification. In subsequent compilations, if CBO generates a verified non-accepted plan (i.e. a previous test plan), an accepted plan is chosen for execution. The accepted plan (previous reference plan) is set as the new test plan, whereas the CBO generated plan is used as the reference plan. This process of verifying accepted plans is called Reverse Verification. Reverse verification provides one more chance for a previous test plan to be accepted in case the performance of the test plan is currently better than the old reference plan.

In cases where the non-accepted unverified test plan performance is worse than average reference plan performance, or the execution times out, during initial test plan performance verification, the reference plan may actually be worse than the test plan with test plan binds and in the test plan environment. Reverse Verification resolves these cases and sets the optimal statuses for the old test plans in the SPM repository in subsequent compilations/executions. Whenever a good test plan is thus missed by the SPM on initial verification, the Reverse Verification module identifies such cases and helps set appropriate plan statuses. If performance of the previously "worse" test plan is better, it is now accepted. Going forward, the optimizer picks a plan with the least optimizer cost among accepted plans.

\subsection{Illustrative Scenarios}
The following examples illustrate how Real-Time SPM behaves in two common situations: bind-sensitive plan choice and a true regression after an upgrade.
\subsubsection{Bind-sensitive plan choice}
Suppose a statement has a previously seen non-accepted plan P1 that is good for a selective bind. On a later compilation with an unselective bind, the optimizer generates P2. Real-Time SPM executes P2 as the test plan and compares it with P1 as the reference plan. If P1’s historical metrics still appear better, P1 is accepted and P2 is marked for reverse verification rather than permanently rejected. If the optimizer later generates P2 again for an unselective bind, reverse verification executes P1 under the current bind and uses P2 as the reference plan. If P2 is then better, it is accepted as well. Thereafter, Real-Time SPM can choose between accepted plans P1 and P2 according to the current binds and optimizer environment. If a later plan P3 appears for another bind, cost-sanity checks and least-cost reference-plan selection are used to choose the most relevant reference plan, and the normal workflow repeats.
\subsubsection{Regression prevention after upgrade}
Suppose a statement used plan P1 before a database upgrade, and after the upgrade the optimizer generates P2. Real-Time SPM executes P2 as the test plan and compares it with P1 as the reference plan. If P1 fails cost-sanity checks and P2 appears better than P1’s historical metrics, Real-Time SPM still does not immediately accept P2; consistent with its conservative preference for previously seen plans when the comparison is uncertain, it instead marks P2 for reverse verification. On the next execution, if the optimizer again generates P2, reverse verification executes P1 in the current environment and uses P2 as the reference plan. If P2 is then found to be worse, P1 is accepted if not already accepted and P2 remains non-accepted. Subsequent executions therefore fall back to P1, avoiding repeated executions of the upgrade-induced regressed P2.

\section{Experimental Results}
We evaluated Real-Time SPM on our Autonomous Database (ADB) cloud fleet, as well as on a customer OLTP workload, and measured its improvement with respect to Auto SPM on a customer data-warehouse workload. The following subsections detail all experiments and their results.

\subsection{Real-Time SPM on ADB}
Real-time SQL plan management logs diagnostic and performance data, such as plan hash values and buffer gets, directly in the data dictionary.

We analyzed data from 20,015 production and test databases running diverse workloads in Oracle Autonomous Databases over five days. No database patches were applied during this time, so the observed regressions were likely due to DML activity, access path changes, and other environmental factors.

\subsubsection{High-level Results}
We observed 959,911 verification events for 463,395 unique SQL statements. Multiple verifications per SQL statement occur because:
\begin{itemize}
  \item Some statements incurred more than one plan change.
  \item Some statements had plans that were re-verified during reverse verification.
\end{itemize}

\indent
Verifications are classified as "normal" and "reverse." A normal verification occurs during the initial interaction with Real-Time SPM when a plan change has occurred. If there is a subsequent reverse verification, this is marked "reverse" in the diagnostic data.

Of 921,820 normal verifications, 731,232 confirmed new plans were similar or better. Out of 190,588 potential regressions, the CBO generated the same plan for 36,017 of them later and went through reverse verification. It accepted the old test plan (and reversed the decision) for 1,380 of them. So, we can say it prevented regressions for 190,588 – 1,380 = 189,208 SQL statements in total. 

Table~\ref{tab:highLevelResults} summarizes these results, showing the breakdown of verifications and their outcomes.

\begin{table}[h]
  \caption{Summary of Results}
  \label{tab:highLevelResults}
  \begin{tabular}{lc}
    \toprule
    Data Point & Value \\
    \midrule
    \verb|Distinct SQL statements|  & 463,395 \\
    \verb|Normal verify: similar performance|  & 426,912 \\
    \verb|Normal verify: better performance|  & 304,320 \\
    \verb|Normal verify: worse performance|  & 190,588 \\
    \verb|Normal verifications | & 921,820 \\
    \verb|Reverse verifications|  & 36,017 \\
    \verb|Reverse verification changes decision |  & 1,380 \\
    \verb|SQL performance regressions prevented|  & 189,208 \\
    \verb|Total verifications (normal and reverse)| & 959,911 \\
    \bottomrule
  \end{tabular}
\end{table}

To assess the magnitude of regressions avoided by Real-Time SPM, we calculated a "regression factor" for each verification as the ratio of test plan buffer gets over reference plan buffer gets. Buffer gets is one of the primary metrics Real-Time SPM uses to measure SQL statement performance. However, SPM also tracks CPU usage to identify cases where fewer buffer gets come at the cost of much higher CPU, which still counts as a regression.

\begin{figure}[h]
  \centering
  \includegraphics[width=\linewidth]{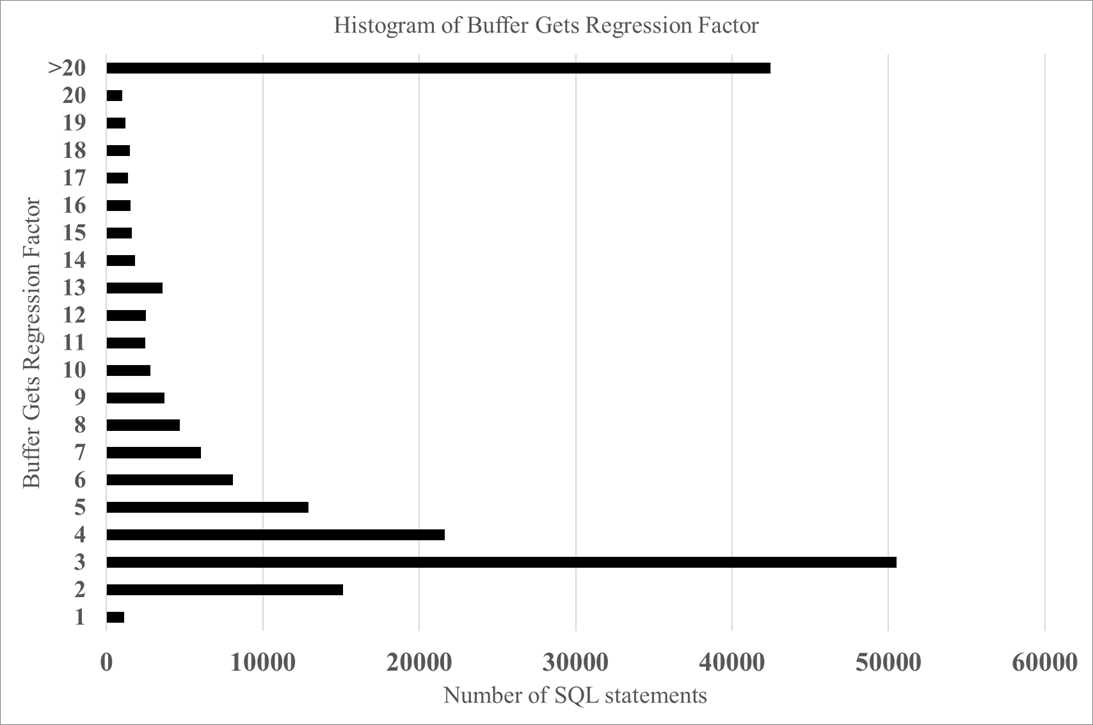}
  \caption{Real-time SPM Buffer Gets Histogram}
  \label{fig:regressionHistogram}
\end{figure}

\autoref{fig:regressionHistogram} shows the distribution of regression factors where the test plan is marked "worse." The peak is at 3, indicating that the majority of test plans exhibit a performance regression characterized by an increase in buffer gets of between two and three times compared to the reference plan.

The histogram shows that 1,153 test plans considered "worse" had fewer buffer gets than the reference plan ("1" in the chart, indicating a regression factor below one). This demonstrates that Real-Time SPM also considers CPU usage when evaluating performance, not just buffer gets.

Some statements regress in buffer gets by more than 20x, and the maximum regression factor is over 6.7 million. Table~\ref{tab:regressionSummary} provides a summary. The high maximum value and large standard deviation indicate a wide variation in the regression factor, and some SQL statements would have suffered significant performance regressions had Real-Time SPM not intervened. This is also reflected in a relatively large mean regression factor compared to the median. The median value is close to the peak shown in \autoref{fig:regressionHistogram}.

\begin{table}[h]
  \caption{Regression Factor Summary}
  \label{tab:regressionSummary}
  \begin{tabular}{lc}
    \toprule
    Regression Factor & Value \\
    \midrule
    \verb|Mean| & 1,800.34 \\
    \verb|Median|  & 3.84 \\
    \verb|Standard Deviation|  & 38,263.02 \\
    \verb|Maximum|  & 6,718,453 \\
    \bottomrule
  \end{tabular}
\end{table}

\subsection{Real-Time SPM vs Auto SPM}
 We devised an experiment to measure the success of both Auto SPM and Real-Time SPM, and highlight the advancements of Real-Time SPM. Real-Time SPM and Auto SPM were executed in a test customer data-warehouse workload with 25 tables (4 of which were partitioned), 2,117 statements and no indexes or DML statements. The query workload was executed 3 times for Auto SPM and 4 times for Real-Time SPM with a specific action (described below) between each run. The experiment was conducted using Oracle Database 26ai release 23.26 - Oracle Enterprise Linux machines with AMD EPYC Processor and 60 CPUs. We used buffer gets and CPU time (in minutes) as the performance metrics. Other metrics like elapsed time/latency varied widely between multiple executions of the same execution plan, and as CPU time is a function of parse time, execution time, and other performance indicators, it - combined with Buffer Gets - is the most stable metric to use.

~\autoref{tab:expRuns} shows the different experimental runs and the action or change prior to each run. The first experiment was for Auto SPM and the second experiment was for Real-Time SPM. 'OFE' stands for Optimizer Features Enable, essentially telling the database to behave like the optimizer of database version X.

\begin{table}[h]
  \caption{Workload Runs}
  \label{tab:expRuns}
  \begin{tabular}{lcll}
    \toprule
    Prior Action & Run & Experiment & Which SPM Enabled \\
    \midrule
    \verb|OFE 18.1| & 1 & Auto SPM & SPM Disabled\\
    \verb|OFE 23.1| & 2 & Auto SPM & Auto on, Real-Time off\\
    \verb|OFE 23.1| & - & Auto SPM & Background SPM task on\\
    \verb|OFE 23.1| & 3 & Auto SPM & No SPM activity\\
    \verb|--------| & - & -------- & -------------------------------- \\
    \verb|OFE 18.1| & 1 & Real-Time & SPM Disabled\\
    \verb|OFE 23.1| & 2 & Real-Time & Real-Time on, Auto off\\
    \verb|OFE 23.1| & 3 (RV) & Real-Time & Real-Time on, Auto off\\
    \verb|OFE 23.1| & 4 & Real-Time & No SPM activity\\
    \bottomrule
  \end{tabular}
\end{table}

The experiments start with OFE 18.1 and then switch to OFE 23.1 in Run 2, thereby inducing plan changes due to an upgrade. We use OFE here as a controlled mechanism to mimic a software upgrade that causes the CBO to generate different plans, some of which may be regressed. For Auto SPM, Run 3 is executed after the background task creates and accepts plan baselines. For Real-Time SPM, Run 3 keeps Real-Time SPM enabled to allow reverse verification (RV) of plans that were verified but remained non-accepted in Run 2, and Run 4 measures workload performance after both initial verification and reverse verification have taken effect.

The 2,117 statements were executed concurrently in 5 batches for both experiments. \autoref{tab:task} displays task duration and the number of accepted plan baselines after each run. 

\begin{table}[h]
  \caption{SPM Task Details}
  \label{tab:task}
  \begin{tabular}{lcc}
    \toprule
    Task & Duration (hours) & Plan Baselines \\
    \midrule
    \verb|Auto SPM Run 2| & 2.35 & NA\\
    \verb|Auto SPM Background Task| & 8 & 63\\
    \verb|Real-Time SPM Run 2| & 2.28 & 153\\
    \verb|Real-Time SPM Run 3 (RV)| & 2.1 & 64\\
    \bottomrule
  \end{tabular}
\end{table}

In our setup, the Auto SPM background task required 8 hours to accept 63 plan baselines. In contrast, Real-Time SPM accepted 153 baselines during Run 2 and another 64 during Run 3, for a total of 217 accepted baselines in about 4 hours and 20 minutes. This indicates that Real-Time SPM both acted on more statements and did so substantially faster than Auto SPM, addressing the background-resource limitations described in Section 2.4.

\begin{figure}[h]
  \centering
  \includegraphics[width=\linewidth]{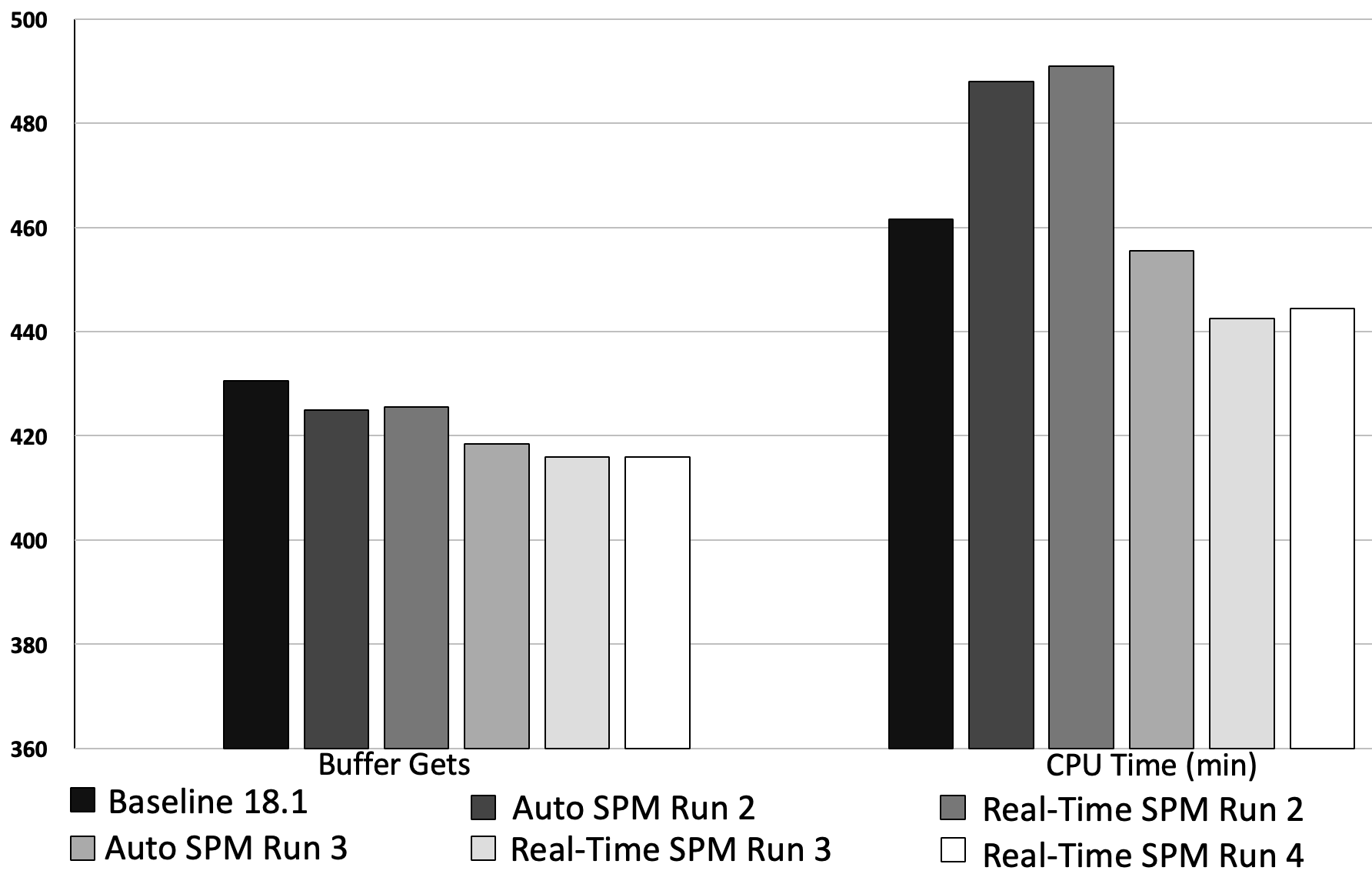}
  \caption{Workload Performance Comparison}
  \label{fig:workloadcomp}
\end{figure}

\autoref{fig:workloadcomp} summarizes the [normalized] workload performance with and without SPM in both experiments, i.e. baseline performance compared to Auto SPM experiment run 3 and Real-Time experiment runs 3 and 4. Auto SPM seems to help in reducing CPU time (462 to 455 minutes), but the effect of Real-Time SPM is more pronounced, as CPU time decreased to about 443 minutes. Buffer Gets also had a reduction in Real-Time SPM runs 3-4 compared to the baseline run. The similar performance metrics of Runs 3 and 4 imply that the plan baselines captured in Run 3 due to reverse verification did not have significant impact on performance. Real-Time SPM run 2 was when all the foreground verification work occurred, and as expected, we see an increase in execution CPU time for run 2, from 462 minutes to about 490 minutes. The increase in execution time for Run 2 in both Auto SPM and Real-Time SPM is due to the new/test plan executions that are suboptimal compared to the baseline run plans.

We also attempted to quantify the compilation overhead introduced by Real-Time SPM. In our experiment, compilation CPU and buffer gets were lower in Runs 2–4 than in Run 1. However, these later runs coincided with an OFE upgrade that includes optimizer and compilation-time improvements. As a result, compile-time comparisons across OFE versions would conflate any Real-Time SPM overhead with unrelated optimizer enhancements.

\begin{figure}[h]
  \centering
  \includegraphics[width=\linewidth]{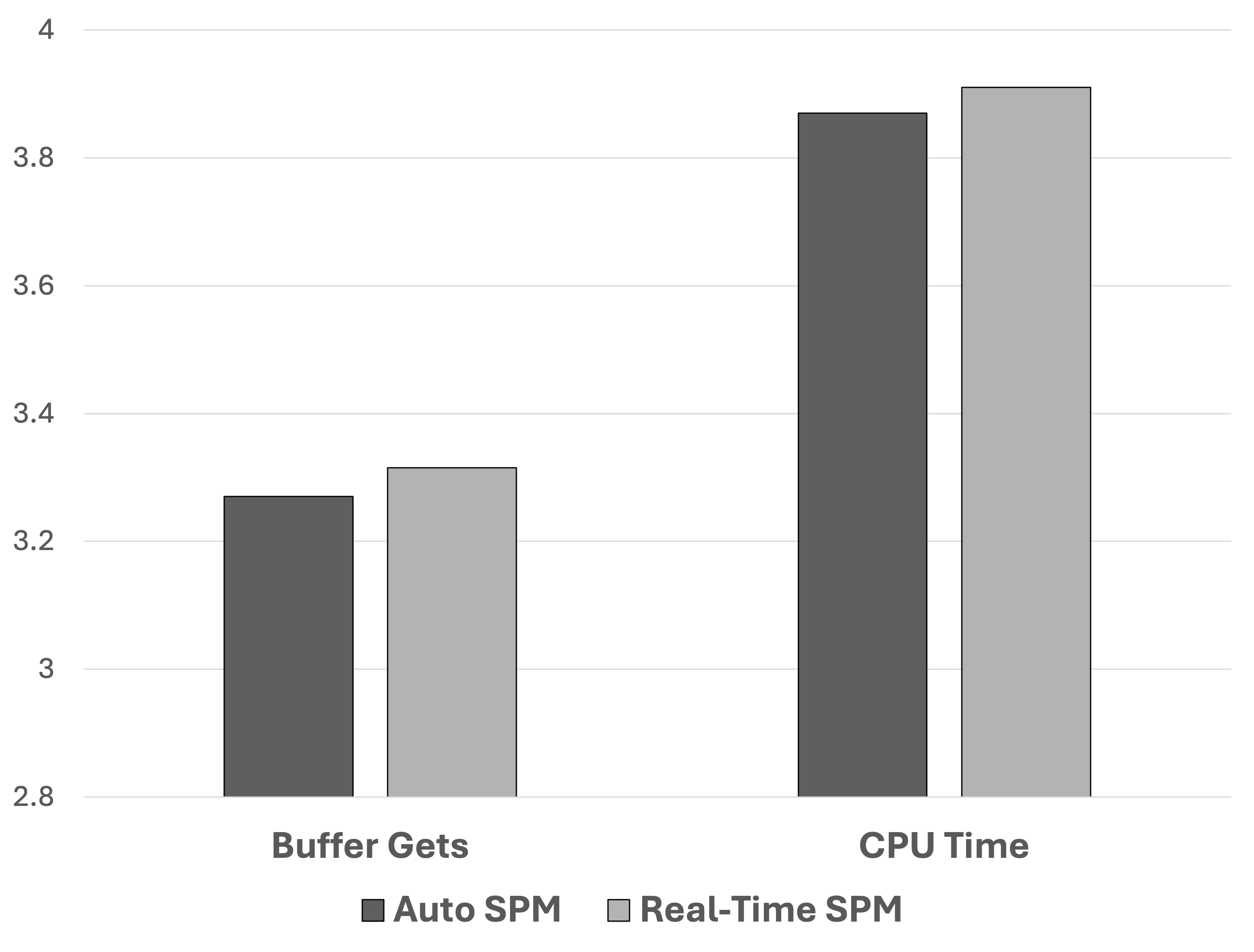}
  \caption{Real-Time SPM Compilation Overhead}
  \label{fig:overhead}
\end{figure}

To isolate the incremental overhead attributable specifically to Real-Time SPM, we compared Run 2 under Auto SPM with Run 2 under Real-Time SPM; both were executed on the same OFE version (23.1). Because Auto SPM operates in the background, Run 2 under Auto SPM effectively serves as the baseline for OFE 23.1 without foreground SPM work, making it a suitable reference for gauging the additional overhead introduced by Real-Time SPM. \autoref{fig:overhead} shows the [normalized] observed overhead.

Real-Time SPM increases compilation buffer gets by roughly 2 KB (about 110 KB with Real-Time SPM vs. 108 KB with Auto SPM). This uptick is negligible compared with overall workload execution buffer gets (more than 1200 MB, which is normalized in \autoref{fig:workloadcomp}) but is consistent with the extra compilation-time work Real-Time SPM performs — like loading historically accepted plan baselines and, when needed, updating plan status in the SPM repository. Compilation CPU time also rises slightly, from 3.87 minutes under Auto SPM to 3.91 minutes under Real-Time SPM. While this is immaterial relative to the total execution time (more than 400 minutes), the overhead is expected given the work required to find and cost reference plans. Our efficient bit-vector logic helps keep this compile-time overhead minimal.

\subsection{Real-Time SPM with Automatic Indexes}
\begin{figure}[h]
  \centering
  \includegraphics[width=\linewidth]{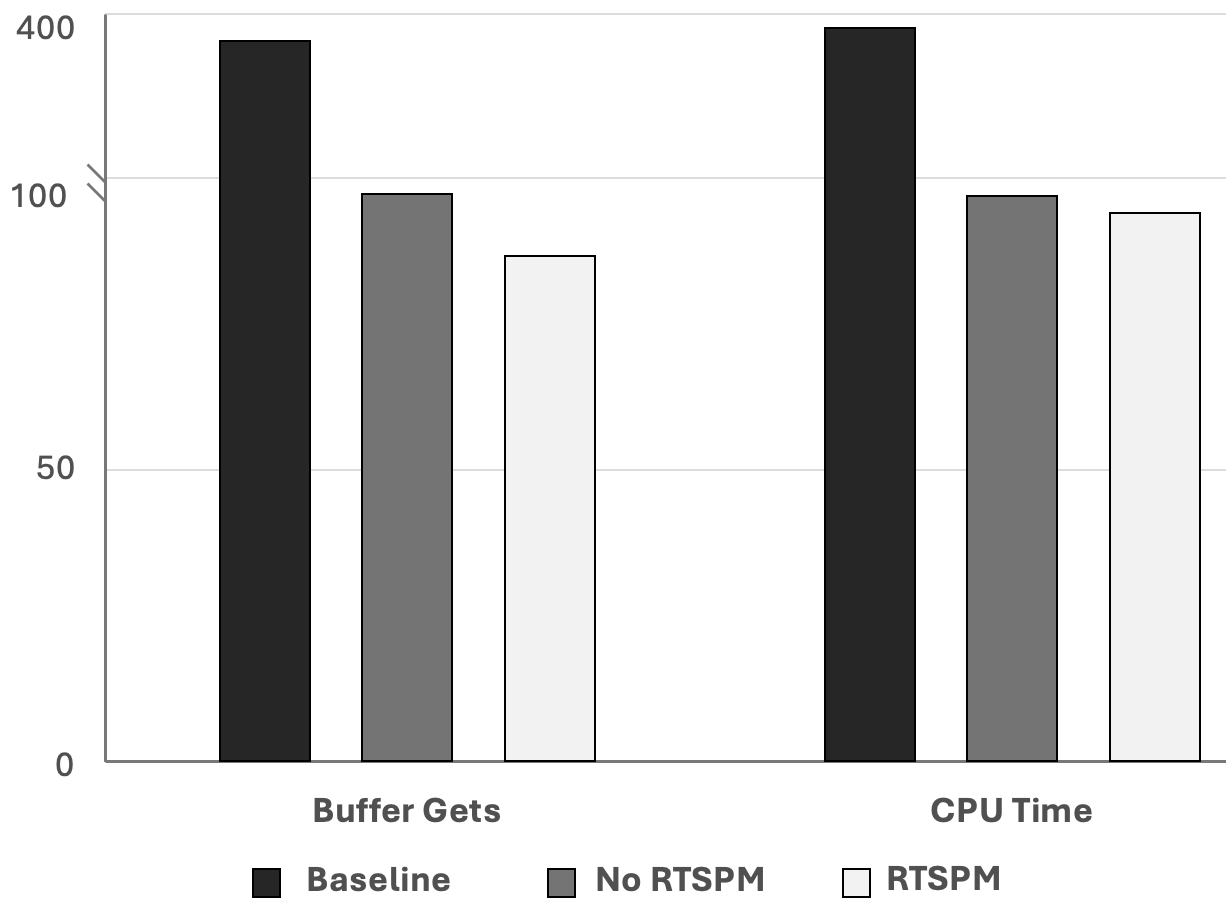}
  \caption{Real-Time SPM with Indexes}
  \label{fig:index}
\end{figure}

We next evaluated Real-Time SPM after introducing indexes into the same customer data-warehouse workload used above. Specifically, we let Automatic Indexing~\cite{Autoindex} create 204 new indexes and then enabled Real-Time SPM to verify the indexed-plans. We then measured workload performance with and without Real-Time SPM enabled, using the same system configuration as in Section 4.2. \autoref{fig:index} shows CPU Time (in minutes) and normalized Buffer Gets for this experiment.

The baseline corresponds to the same workload before Automatic Indexing created any indexes and without any SPM activity. The 4x improvement from the baseline to the indexed runs, both with and without Real-Time SPM, implies that the new indexes substantially improved overall workload performance. At the same time, some individual statements still regressed (not pictured above), with the largest observed regressions reaching about 100x in Buffer Gets and about 3x in CPU Time. Relative to the configuration without Real-Time SPM, Real-Time SPM reduced Buffer Gets by about 11\% and CPU Time by about 3\% for the entire workload, indicating that it helped mitigate regressions caused by index-driven plan changes while still preserving the benefits of the new access paths.

\subsection{Real-Time SPM in OLTP}
We validated Real-Time SPM on an OLTP workload with 39,000 queries as well. This time, we downgraded from a higher OFE to a lower OFE, where regressions might be more likely. The setup details are described in \autoref{tab:oltpRuns}. 

\begin{table}[h]
  \caption{OLTP Workload Runs}
  \label{tab:oltpRuns}
  \begin{tabular}{lcll}
    \toprule
    Prior Action & Run & Which SPM Enabled \\
    \midrule
    \verb|OFE 22.1| & 1 & No SPM\\
    \verb|OFE 18.1| & 2 & Real-Time on\\
    \verb|OFE 18.1| & 3 & Real-Time on (reverse verification)\\
    \verb|OFE 18.1| & 4 & No SPM\\
    \bottomrule
  \end{tabular}
\end{table}

The results are summarized in \autoref{tab:oltpPerf}. Buffer Gets (megabytes) and CPU Time (minutes) were analyzed comprehensively as well as just during compilation to measure any discernible overhead of Real-Time SPM.

\begin{table}[h]
  \caption{OLTP Workload Performance}
  \label{tab:oltpPerf}
  \begin{tabular}{cccccc}
    \toprule
    Run & BG (Total) & BG (Comp.) & CPU (Total) & CPU (Comp.)\\
    \midrule
    1 & 280 & 2.68 & 170.37 & 35.50 \\
    2 & 307 & 2.64 & 206.83 & 31.28 \\
    3 & 276 & 2.60 & 171.72 & 29.08 \\
    4 & 274 & 2.60 & 167.95 & 28.98 \\
    \bottomrule
  \end{tabular}
\end{table}

In Run 2, total buffer gets and CPU time increased as expected due to the OFE downgrade and the resulting plan degradation. Compilation-related metrics nevertheless decreased while Real-Time SPM was active, which we attribute to confounding effects from the OFE change (i.e., OFE-dependent compilation behavior improvements), rather than to Real-Time SPM reducing compilation cost. Runs 3 and 4 then demonstrate the effectiveness of Real-Time SPM: buffer gets and CPU time return to the Run 1 baseline and show a modest additional improvement. Overall, these results indicate that Real-Time SPM can preserve plan stability while still enabling plan evolution by leveraging both historical baselines and newly generated plans, delivering sustained performance benefits in subsequent executions with minimal compilation overhead.

\section{Related Work}
Query performance regressions caused by execution plan changes are a recurring operational pain point across OLTP, analytics, and mixed workloads. This section positions Real-Time SPM relative to mechanisms that exist in production today across major cloud and enterprise database platforms, as well as approaches suggested through research.

Several production database systems provide functionality related to query plan stability, though they differ substantially in scope and guarantees. Amazon Web Services Aurora PostgreSQL ~\cite{QPM} offers plan management features such as plan hints and limited plan pinning to capture and force “good” plans, but this is largely manual and lacks automated runtime regression detection or rollback. Google Cloud Spanner ~\cite{spanner} supports optimizer plan and optimizer-version pinning, allowing users to freeze plan behavior across upgrades; however, this approach is coarse-grained, reactive, and does not adapt to workload or data evolution at runtime - limiting the scope for performance gains through new access paths. Microsoft SQL Server Query Store ~\cite{queryStore} provides intelligent plan forcing integrated with automatic plan correction. If a forced plan degrades, Query Store can "unforce" it. The optimizer is allowed to pick a better plan if the forced one fails quality thresholds, but only regressions in the forced plan are detected. It is more reactive and relies more on manual intervention or offline analysis rather than automatic, proactive real-time regression prevention. IBM Db2 ~\cite{db2}  can persist the prepared form (runtime structures) of a dynamic SQL statement in the catalog. On subsequent executions, instead of fully re-preparing (and potentially choosing a different access path), Db2 can reload the saved prepared runtime structures, preserving the previous access path and behavior. Db2’s dynamic plan stability focuses on preserving the last prepared access path for dynamic SQL, but does not maintain a catalog of multiple accepted plans nor does it provide verified regression guarantees like Real-Time SPM does. Manual identification and stabilization is also required. PingCAP TiDB SQL bindings ~\cite{TiDB} allow users to bind queries to specific plans or hints, offering deterministic control but requiring manual specification and ongoing maintenance. 

In contrast, Oracle Real-Time SPM uniquely combines automatic regression detection, historical plan baselines, and deterministic plan verification and selection at runtime. It enables conservative, workload-wide plan management without requiring user intervention or prior plan pinning, while also allowing for improved plans.

Recent studies in plan stability intend to hedge against cardinality estimation errors. Robust Query Optimization via Parallel Multi-Plan Execution \cite{Rome} (ROME) addresses query performance regressions caused by cardinality estimation errors through a runtime multi-plan execution strategy. ROME generates a small set of alternative plans by modifying optimizer features and executes them in parallel, terminating once the first plan completes. A cost-aware probabilistic model is used to limit the plan search space and minimize expected execution cost, providing a statistical guarantee against severe mis-optimization. However, ROME does not retain knowledge across executions: candidate plans are regenerated for each invocation and discarded thereafter, and the guarantee is probabilistic rather than deterministic. In contrast, Oracle Real-Time SPM accumulates evidence over historical executions and enables plan stability over time, ensuring performance no worse than a verified baseline. While ROME can obtain a good plan even on the first execution of a query and hedge against extreme cardinality errors, SPM is general and not just targeting cardinality mis-estimates. Accordingly, ROME and Real-Time SPM are best viewed as orthogonal approaches.

Alternative approaches to mitigating query performance regressions include parametric query optimization (PQO) ~\cite{par2q0, APQO, SPQO} and adaptive query execution (AQE) like in Databricks ~\cite{databtricks, HTAP} and as referenced in the beginning of section 1 as well. PQO addresses the challenge of repeatedly optimizing the same SQL template under varying parameter bindings. Oracle addresses this problem through Adaptive Cursor Sharing, or automatically deciding whether an existing plan is likely to be optimal for a new bind value ~\cite{cardFeed}. Earlier learned PQO techniques typically partition the parameter space and associate each region with a fixed set of cached plans, which limits their ability to adapt when workload distributions shift or when new execution plans become necessary. More recent advances ~\cite{APQO} model execution cost as a function of both query parameters and candidate plans, enabling plan comparison even as the plan cache grows dynamically. This design supports fine-grained, per-execution plan selection under changing workloads while retaining a safe fallback to the underlying optimizer. While effective for low-dimensional parameter spaces, with most of the cost incurred at compile time and little to no runtime overhead, this strategy suffers from parameter explosion as the number of predicates and parameter bindings grows, leading to an exponential increase in regions and candidate plans. AQE, in contrast, focuses on correcting execution-time mistakes by re-optimizing queries based on statistics collected during execution ~\cite{databricks_online}. It executes part of the query, collects real statistics, then re-optimizes the remaining stages and continues execution with a modified plan. By leveraging runtime feedback, AQE can adjust join strategies, reorder operators, or rebalance execution in the presence of data skew or inaccurate cardinality estimates, thereby improving robustness within a single query execution.

While PQO and AQE both reduce performance degradation, they operate at different layers of the optimization stack than Oracle Real-Time SPM. PQO emphasizes fine-grained adaptivity for bind-sensitive queries, and AQE focuses on intra-query correction during execution, whereas Real-Time SPM enforces conservative plan management at the workload level by detecting regressions and restoring previously verified plans. Furthermore, PQO and AQE are iterative approaches to potentially finding a \textit{new} plan. Oracle itself has adaptive statistics techniques to address cardinality issues \cite{adaptStat}, along with iterative approaches such as adaptive query processing \cite{oracleAQP, oracleAdapt} for intra-query correction and statistics feedback \cite{cardFeed, oracleCF} for post-query correction. Once a new plan is found, Real-Time SPM reasons about historical plan quality across executions and prevents regressions caused by persistent optimizer plan changes. As a result, these approaches are best viewed as complementary: PQO and AQE improve adaptivity and resilience within individual queries, while Real-Time SPM ensures stability across evolving workloads, data distributions, and system changes.

Similar to Real-Time SPM, Microsoft Research’s Plan Stitch ~\cite{planStitch} aims to achieve stability and low-risk plan improvement using historical execution information. Rather than relying solely on "reversion-based plan correction" — where previously observed whole plans are retained and reinstated upon regression — Plan Stitch exploits operator-level execution statistics to recombine efficient subplans drawn from multiple prior executions of the same query. It constructs an AND-OR graph of equivalent subplan groups and applies dynamic programming to select a minimum-cost stitched plan within the space of previously observed operators. This approach can generate a new plan that has not been executed before, yet is composed entirely of historically observed operators, thereby attempting to balance safety and optimization opportunity beyond whole-plan selection. However, Plan Stitch estimates stitched-plan cost by composing previously observed operator costs, making it sensitive to cost feedback staleness under data changes. In contrast, Real-Time SPM mitigates risk through empirical foreground verification under current runtime conditions before admitting a new plan, and, to a large extent, does not rely on any cost model - which might be faulty for the particular query and the very reason a regression is seen. The two approaches are therefore complementary: Real-Time SPM offers a production-safe framework for admitting plan changes, whereas Plan Stitch again expands the space of candidate plans that can be considered safely.

Machine learning techniques have also been proposed to guide the optimizer towards better execution plans. Recent work on learned query optimizers (LQOs) increasingly emphasizes augmenting, rather than replacing, classical optimizers by constraining the search space \cite{perfGuard, pilotScope, autoSteer, genJoin}. Approaches range from improving plan discovery and selection \cite{autoSteer} to learning subplan hints that guide the optimizer toward more promising regions of the plan space \cite{genJoin}. While these techniques can outperform classical optimizers when considering execution time in isolation, recent evaluation studies show that learned optimizers often incur substantial inference overhead and exhibit unstable end-to-end performance \cite{perfGuard}. Moreover, LQOs are susceptible to the cold start problem, where insufficient training data or limited execution history leads to unreliable early decisions, and they lack formal mechanisms to prevent regressions on a per-query or workload-wide basis. These limitations underscore the importance of lightweight, production-grade mechanisms such as Real-Time SPM for robust regression control.

\section{Conclusion}
Execution plans for SQL queries have a significant impact on overall database performance. A variety of factors — including statistics refreshes, data evolution, and optimizer changes — can lead to the selection of new execution plans. While most plan changes are beneficial or benign, a small number can cause severe performance regressions and threaten system stability. Oracle Real-Time SPM addresses this challenge within the optimizer and runtime execution loop by detecting regressions attributable to plan changes during statement execution. When sufficient evidence indicates that a previously observed plan is superior, Real-Time SPM reinstates that plan using SQL plan baselines; conversely, when a new plan demonstrably outperforms existing baselines, it is verified and retained. By enabling plan admissibility through baselines and reliably preventing regressions at runtime, Real-Time SPM addresses a critical dimension of query optimization — long-term stability and predictability in production environments — that remains largely unaddressed by adaptive or generative learned optimization techniques.

We detail the novel architecture of Real-Time SPM, contrasting foreground and background verification of new execution plans. We experimentally demonstrate that foreground verification immediately detects and resolves regressions upon initial query execution, enabling faster and more reliable performance assurance across both cloud-native and on-premise environments.

\begin{acks}
We would like to thank \c{C}etin \"{O}zb\"{u}t\"{u}n for contributing to the novel ideas and algorithmic design of Real-Time SPM. We also extend our appreciation to present and former members of the Oracle Optimizer team including Zhan Li, Palash Sharma, and Iris Zhang; as well as the Auto Task and SQL Tuning Set teams for their contributions to respective aspects of the product life cycle. We also thank Oracle System Testing team for conducting the experiments and helping us measure the efficacy and robustness of Real-Time SPM.
\end{acks}

%\clearpage

\balance

\bibliographystyle{ACM-Reference-Format}
\bibliography{sample}

\end{document}